\documentclass[aps, prl, twocolumn,  amsmath]{revtex4-1}

\usepackage{graphicx}
\usepackage{bm}

\begin{document}

\title{ Attosecond bunches of gamma photons and positrons generated in nanorod array targets}

\author{Zs. L\'ecz}
\email{zsolt.lecz@eli-alps.hu}
\affiliation{ELI-ALPS, ELI-HU NKft. Dugonics square 13., 6720 Szeged, Hungary}

\author{A. Andreev}

\affiliation{Max-Born Institute, Berlin, Germany}
\affiliation{ELI-ALPS, ELI-HU NKft. Dugonics square 13., 6720 Szeged, Hungary}

\date{\today}

\begin{abstract} 
The sub-atomic experimental exploration of physical processes on extremely short time scales has become possible by the generation of high quality electron bunches and x-ray pulses with sub-femtosecond durations. Temporal coherence is crucial for achieving high resolution in diffraction-based diagnostics and if the transverse spatial coherence is also provided, then phenomena with very small cross sections could also be observed. Increasing the photon energy from x-ray to gamma-ray regime makes probing of extremely small space-time domains accessible. Here, a mechanism for generating attosecond gamma photon and positron bunches with small divergence using laser intensities below $10^{23}$ W/cm$^2$ is proposed. Numerical simulations are used to formulate the conditions for coherent radiation and to characterize the generated photon and positron bunches.

\end{abstract}

\maketitle

The generation of short wavelength radiation on the surface of flat metallic foils irradiated by intense laser pulses is one of the fundamental research areas in plasma physics of the 21$^{\text{th}}$ century.
The surface high harmonic generation (SHHG) technique is based on the laser-driven relativistic motion of radiating electrons oscillating near the laser-plasma interface \cite{shortpulse, theorySHHG, Mikhailova, plasmamirror}. In the case of  a close-to-normal incidence, a thin layer of electrons, originating from the plasma skin-depth, oscillates with relativistic velocity due to the ponderomotive force of the incident laser pulse hence this mechanism is often referred to as the relativistic oscillating mirror (ROM). 
Numerical simulations have shown that at larger incidence angles, electrons can leave the plasma surface and are compressed into nano-bunches, which emit coherent synchrotron radiation \cite{andreev, pop_pukhov, attobunches, attospiral}. The maximum energy of photons emitted in these processes is usually at the level of few 100 eV, which corresponds to a wavelength still above the cell size of the grid used in high resolution simulations. Increasing the intensity of laser pulses results in the wavelength of emitted radiation becoming too short for a classical numerical treatment and, thus, a quantized photon emission has to be considered \cite{extendedPIC, SkinDepthEmiss}. This extension of the kinetic simulations becomes important when the laser intensity exceeds $I_L>10^{22}$ W/cm$^2$.

This work uses the EPOCH particle-in-cell code \cite{epoch} which incorporates the Monte-Carlo algorithms, responsible for sampling the quantum-synchrotron spectrum, and the radiation reaction force in each time step \cite{Duclous, Kirk}. The probability of photon emission depends on the parameter $\eta=\gamma E_{\perp}/E_{crit}$, where $\gamma$ is the Lorentz factor of electrons, $E_{\perp}$ is the electric field perpendicular to the electrons momentum and $E_{crit}\approx 1.3\times 10^{18}$ V/m is the Schwinger field limit. The Breit-Wheeler process is implemented for electron-positron pair creation which occurs as the gamma photons propagate in the laser field.
The probability of photon decay is governed by the quantum parameter $\chi = (E_{\perp}/E_{crit})\epsilon_{ph}/(2m_ec^2)$, where $\epsilon_{ph}=h\nu$ is the photon energy. Pair creation becomes significant for $\chi>0.1$ and in these simulations $E_{\perp}/E_{crit}\approx 10^{-3}$ which means that photons with energy greater than $\sim$100 MeV will contribute to pair creation. According to ponderomotive scaling\cite{ponderoScaling} $\epsilon_e=(\sqrt{1+a_0^2/2}-1)m_ec^2$, where $a_0=eE_L/m_e\omega_0c$ is the normalized laser field amplitude with $\omega_0$  the laser frequency, the average electron energy achieved during laser-solid interaction is about 100 MeV for an intensity of $I_L=10^{23}$ W/cm$^2$. 
This is not enough for emission of photons with $\epsilon_{ph}>100$ MeV. 

\begin{figure}[h]
\centering
\includegraphics[width=0.23\textwidth, trim= 10mm 5mm 0 0]{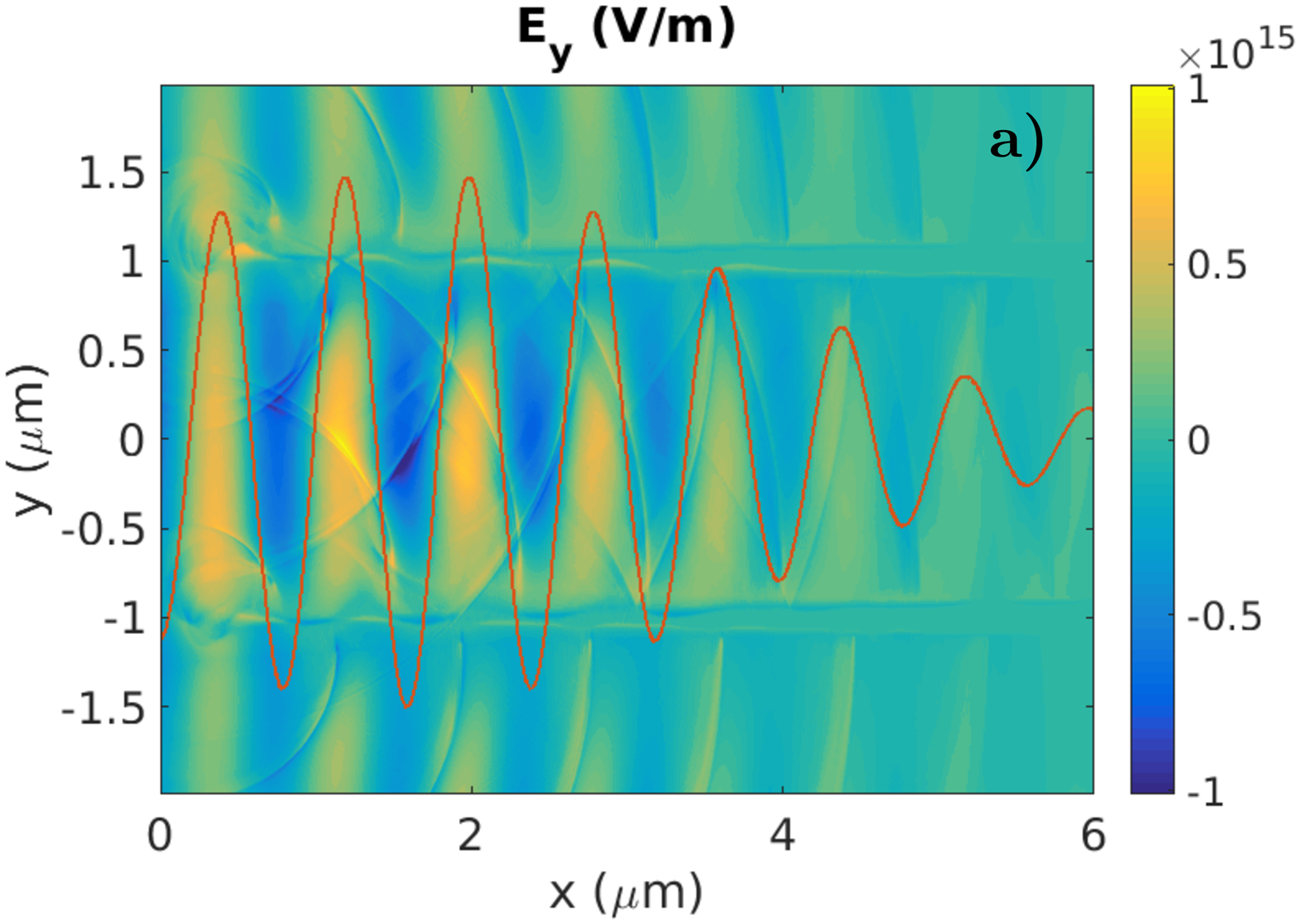}
\includegraphics[width=0.23\textwidth, trim= 10mm 5mm 0 0]{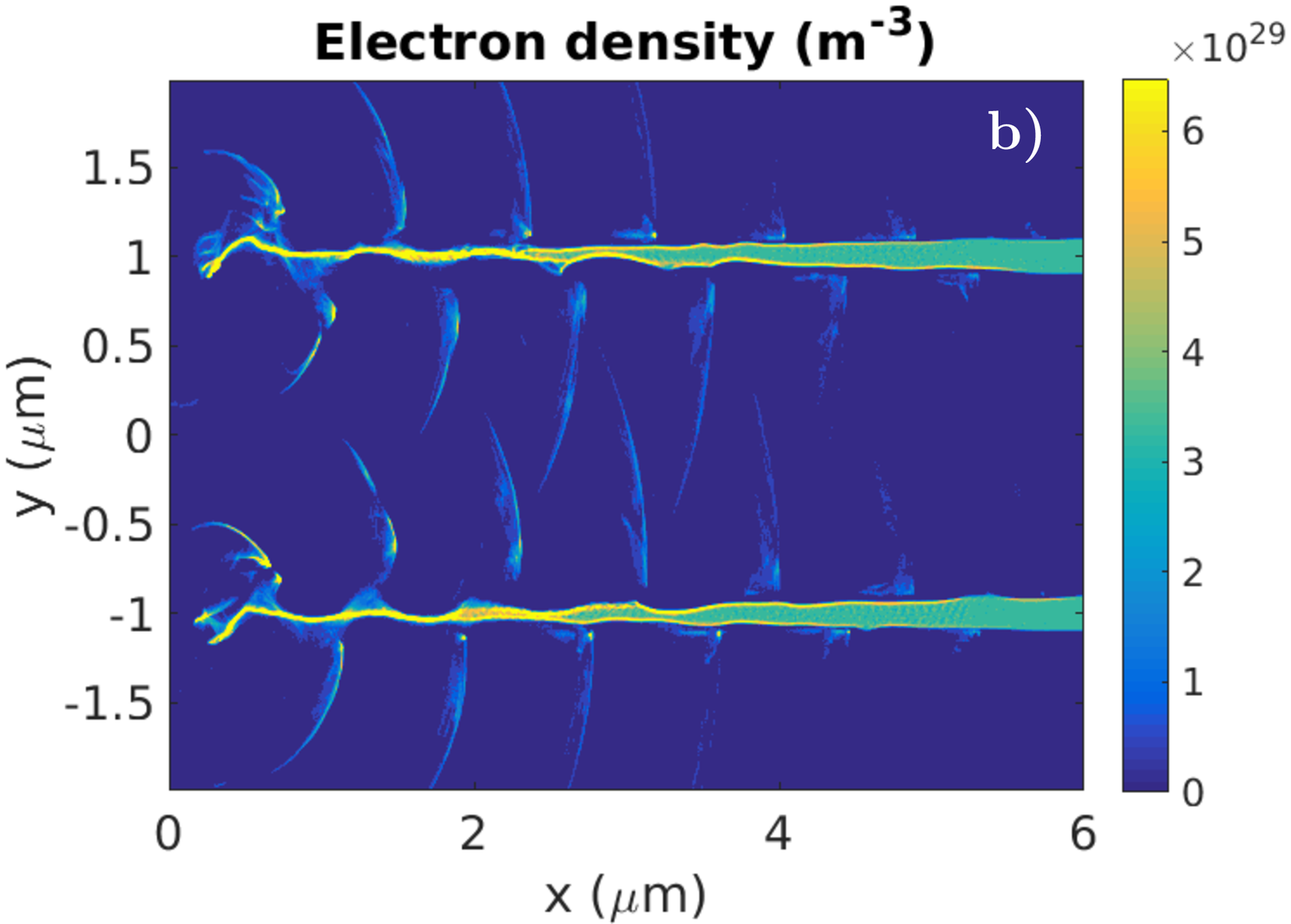}

\caption{ Laser electric field (P-polarized) (a) and electron density distribution (b) at the beginning of the interaction ($t=24$ fs). The laser spot diameter is $2\,\mu$m. }
\label{interaction}
\end{figure}

\begin{figure*}[ht]
\centering
\includegraphics[width=0.31\textwidth]{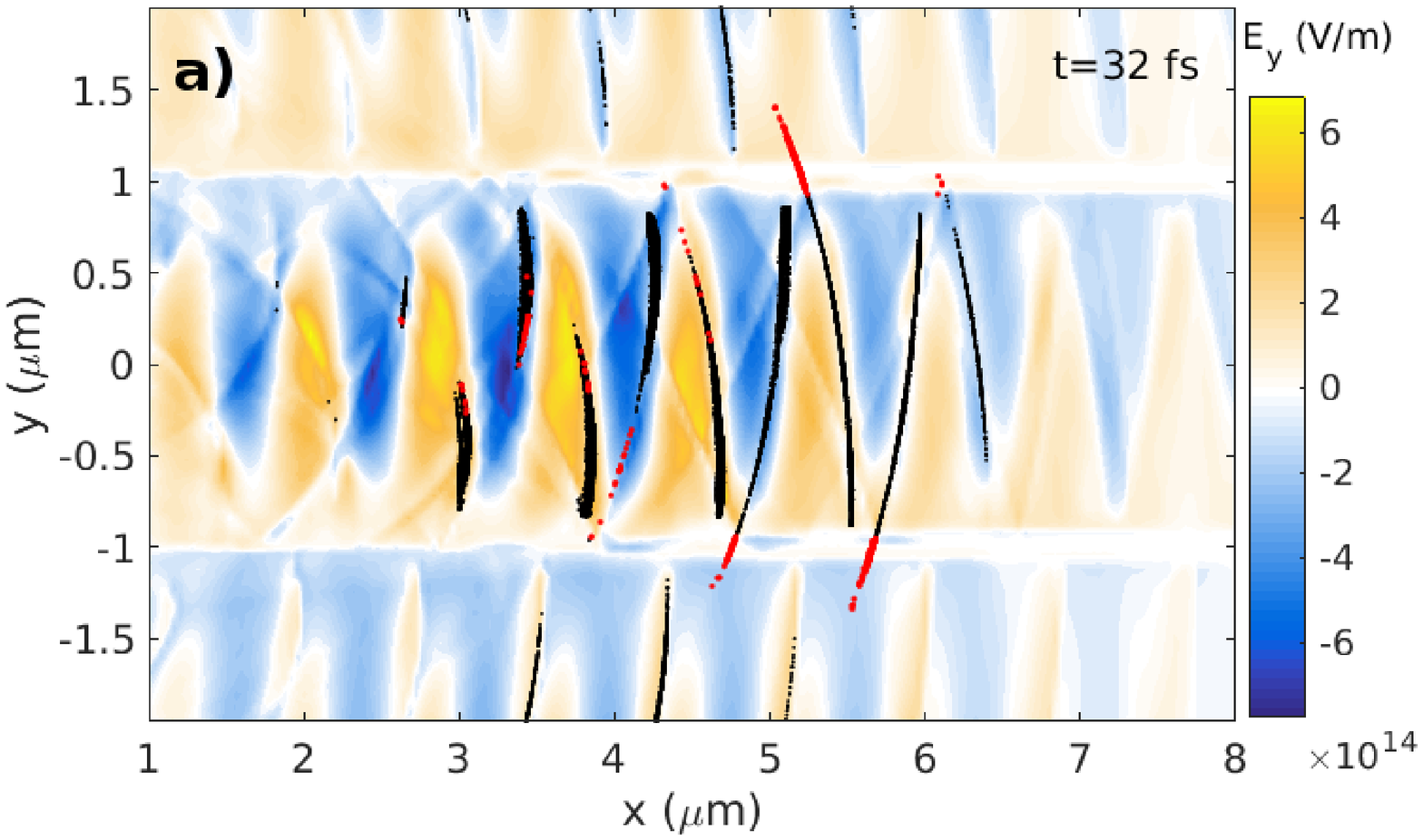}
\includegraphics[width=0.31\textwidth]{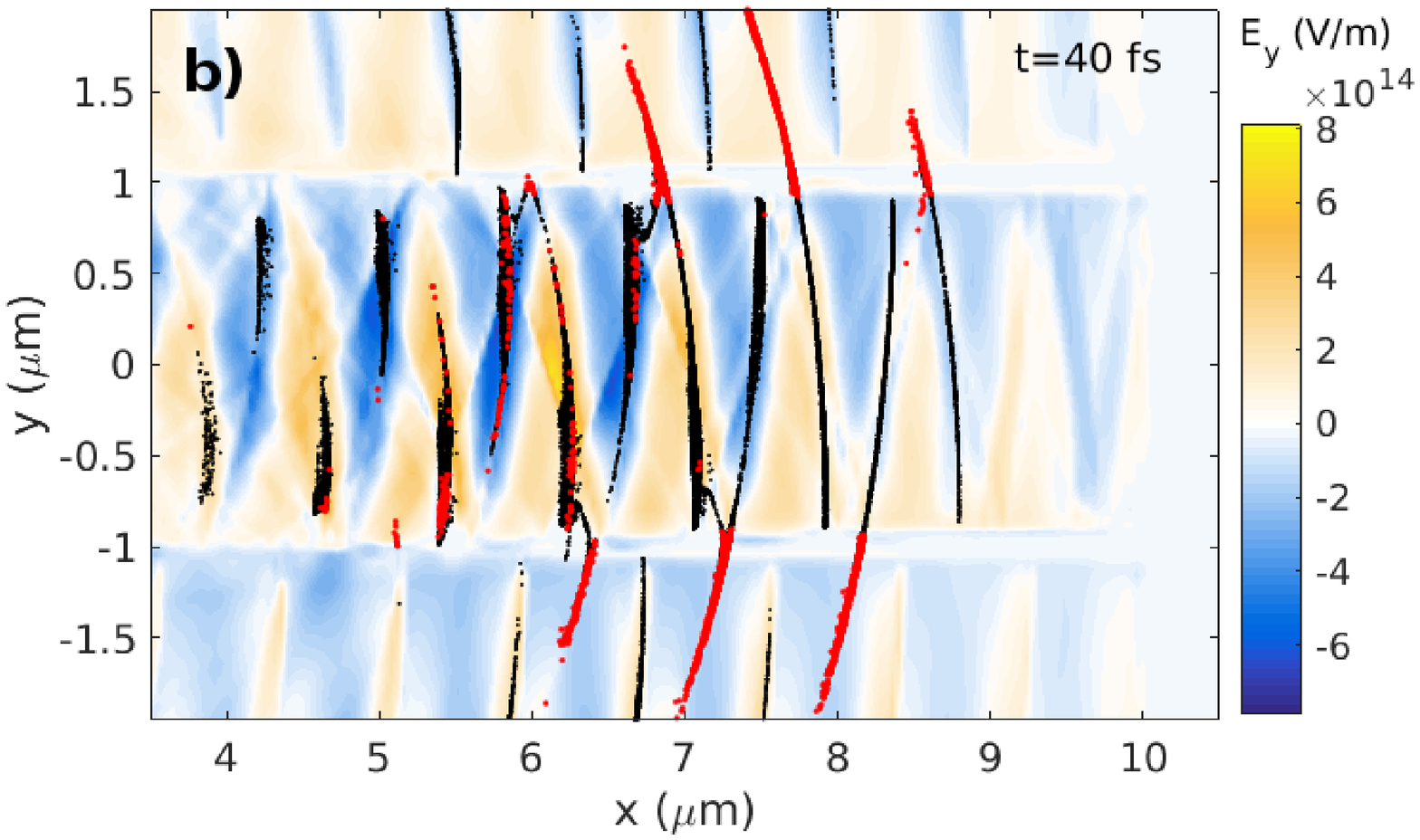}
\includegraphics[width=0.31\textwidth]{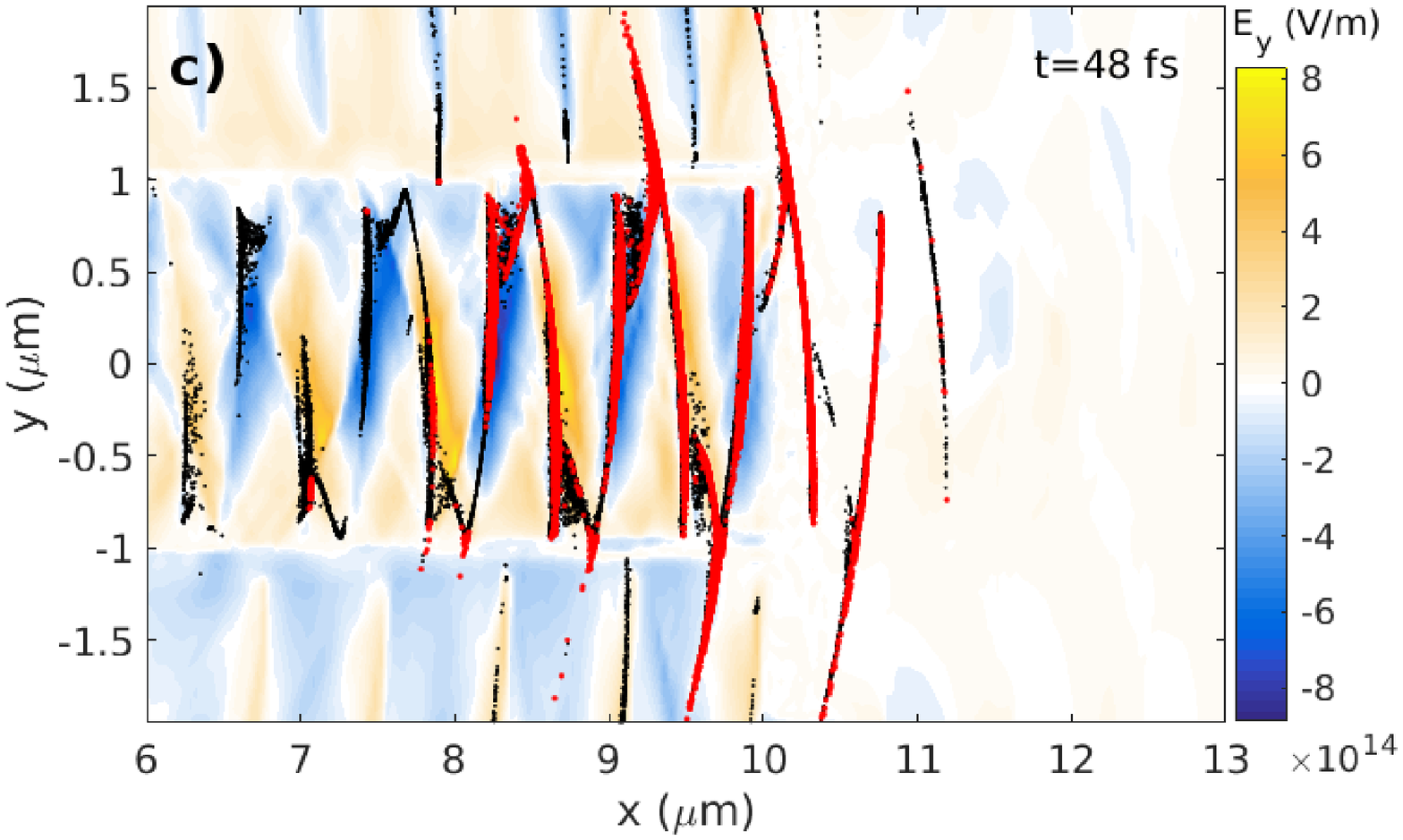}

\caption{ The color scale shows the background electric field. Electrons with energy higher than 100 MeV are shown by black dots, while photons with $\epsilon_{ph}>200$ MeV are shown in (a), (b) and with $\epsilon_{ph}>300$ MeV in (c) by red dots. }
\label{interaction2}
\end{figure*}

Electron acceleration can be significantly enhanced by using nanorod array targets \cite{PukhovNanoR, Andreev2014}, which in turn results in higher photon energies. 
So far, only incoherent radiation has been considered  \cite{nanoforest, PhotonsNanorods} which is a consequent of the chaotic motion of electrons between the closely placed nanorods. At very high intensities, even in the case of plane targets, only incoherent radiation has been discussed \cite{SkinDepthEmiss, NCDplasma}. Here, we show that a suitable large separation ($D_{sp}$) between nanorods allows the formation of attosecond electron bunches at every half cycle of the laser field, resulting in the coherent emission of gamma photons in attosecond bunches.

This works builds on previous works \cite{nanoforest} \cite{microWire} and analyses the interaction of intense laser pulses ($10^{22}$ W/cm$^2 <I_L<10^{23}$ W/cm$^2$) with array of nanorods at solid density including the above mentioned QED effects. The laser pulses considered have a FWHM duration of 15 fs while the density of nanorods is 200$n_{cr}$, where $n_{cr}=\omega_0^2m_e\epsilon_0/e^2$ is the critical density, with a diameter and length of 200 nm and 10 $\mu$m, respectively. A thin foil (0.5 $\mu$m thick overdense plasma slab) is placed at the end of the nanorods, which will reflect the laser pulse. The separation distance between nanorods is an important parameter because it defines the photon emission triggered by the accelerated electrons in the vacuum region i.e., the space between nanorods \cite{alexXray, PhotonsNanorods, transitionNanoW}. Since the aim of this work to produce intense gamma photon bunches with sub-femtosecond duration, the coherency of accelerated electrons and of the subsequent radiation has to be preserved. This is achieved by choosing a separation distance larger than the electron excursion radius, $D_{sp}\gg r_e\approx eE_L/(\gamma m_e\omega_0^2)<\lambda_L$ and thus avoiding the early crossing of fast electrons with the neighboring nanorods. In these simulations $D_{sp}$ is taken to be equal to the laser spot radius, which ensures unperturbed acceleration of electrons. Due to the large vacuum region, the electrons have enough time to reach GeV energy, with only a small fraction reaching the neighboring nanorod. 
 The emitted gamma photons interact with the counter-propagating fields of the reflected pulse and a number decay into $e^-$,  $e^+$ pairs. 

A snapshot from the beginning of the simulation is presented in Fig. \ref{interaction} which shows the laser electric field and the electron density distributions. The simulation domain is represented by $4000\times 1000$ grid points. The peak intensity of the Gaussian laser pulse is $I_L=6\times 10^{22}$ W/cm$^2$. Electron nano-bunches are extracted from the thin overdense rod at each half laser cycle, when the ponderomotive force pushes the electrons forward but consecutive electron bunches are below and above a single nanorod, as also shown in Ref. \cite{microWire}. At the beginning of the interaction, the electron bunches emit EM radiation in the classical limit in a spectral range which can be resolved by the numerical grid used here. Due to the short duration of the bunches, the radiation is coherent and attopulses are produced propagating with a spherical pattern as shown in the upper picture of Fig. \ref{interaction}. The spherical propagation is the result of a diffraction effect as the tip of the nanorod behaves as a secondary point-like light source according to the Huygens-Fresnel principle. Later, the extracted electrons are further accelerated in the laser field between the nanorods (vacuum region) and emit photons with shorter wavelength which is unresolved by the grid. In this phase of the interaction, the QED modules start to play a role which include the emission of gamma photons and their recoil effect on the emitting electron. At this intensity the recoil effect is relatively small and the coherence of further emitted photons is preserved. 

\begin{figure}[h]
\centering

\includegraphics[width=0.3\textwidth]{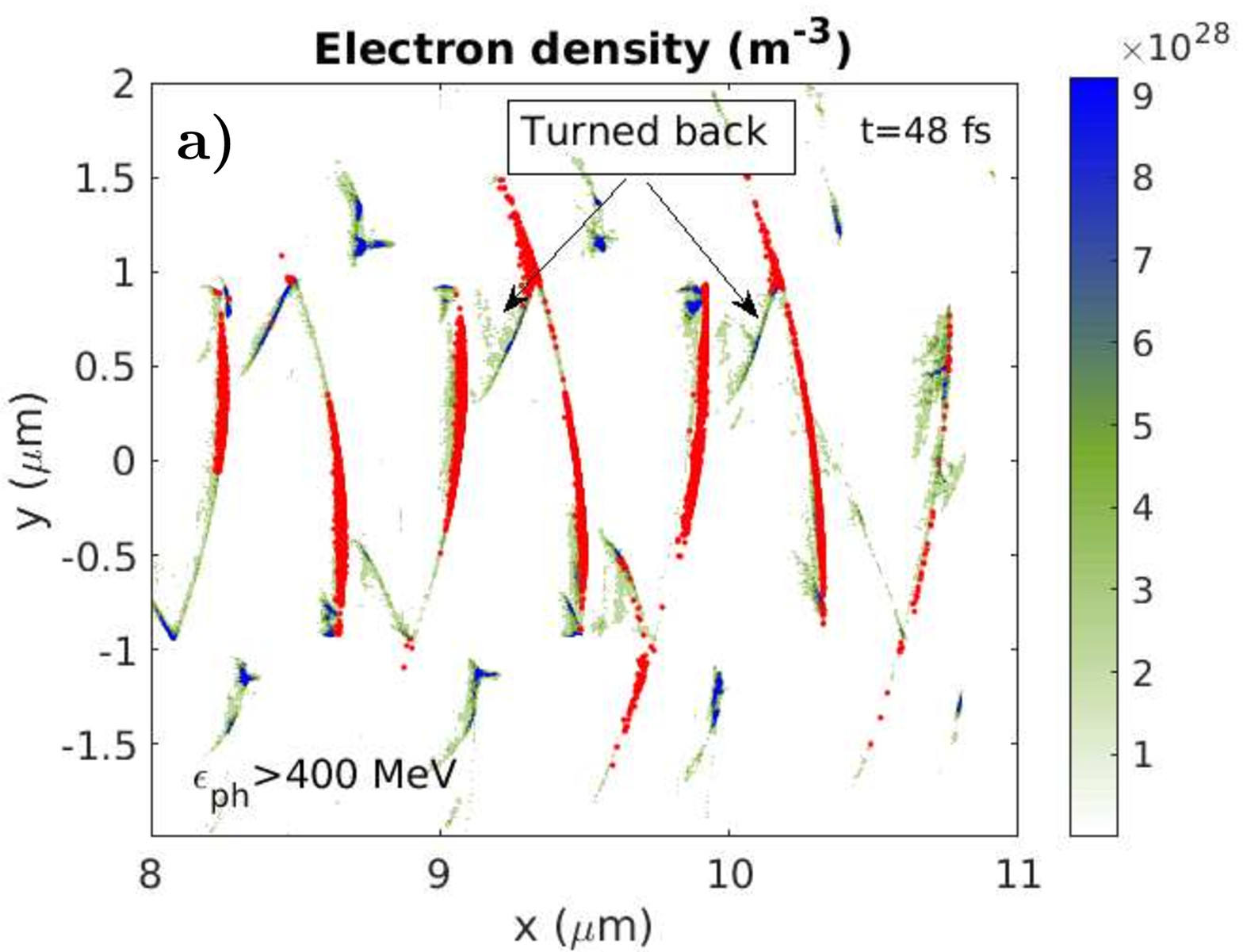}
\includegraphics[width=0.15\textwidth]{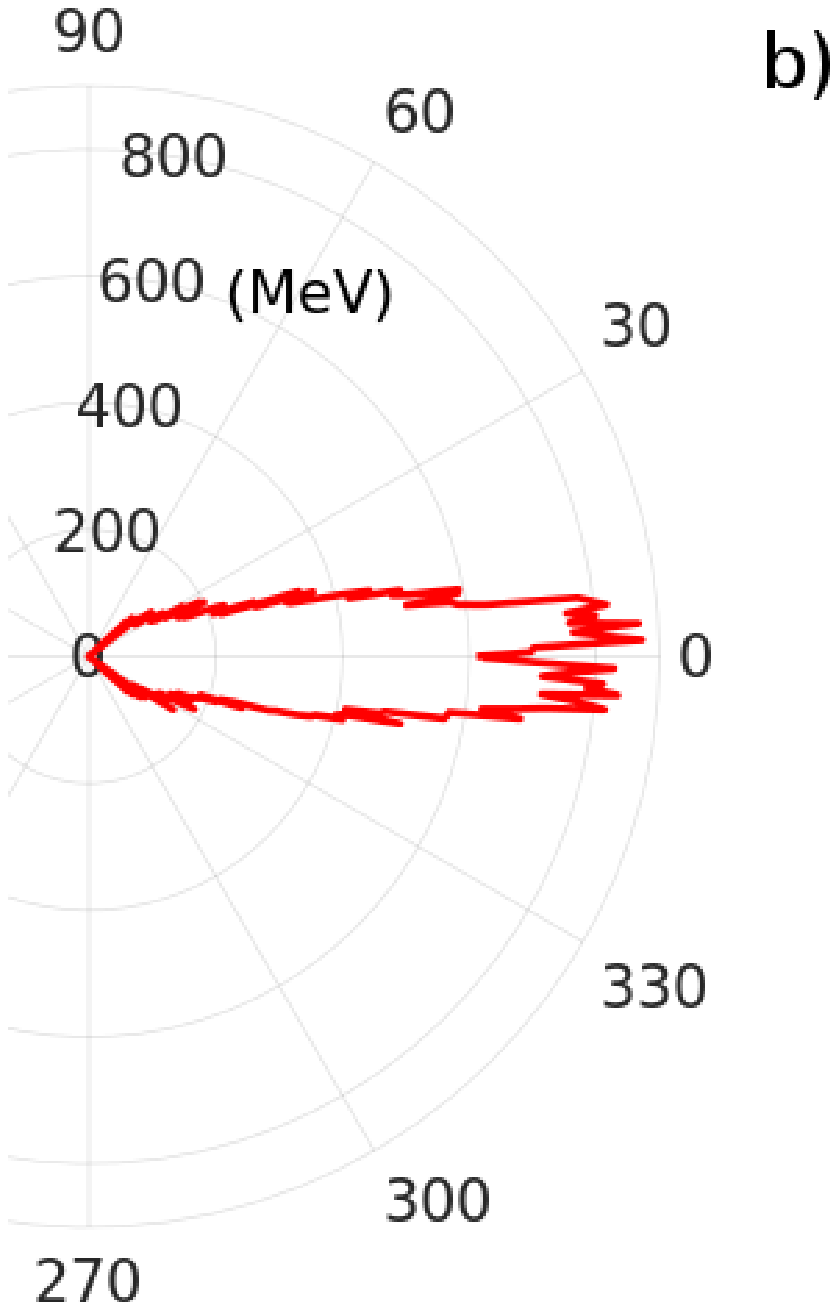}

\caption{ (a) Zoom on the electron density (color scale) at $t=48$ fs and photons are shown with red dots ($\epsilon_{ph}>400$ MeV). (b) Angular-energy distribution of all photons at the same time instance. }
\label{turnBack}
\end{figure}

\begin{figure*}[ht]
\centering

\includegraphics[width=54mm]{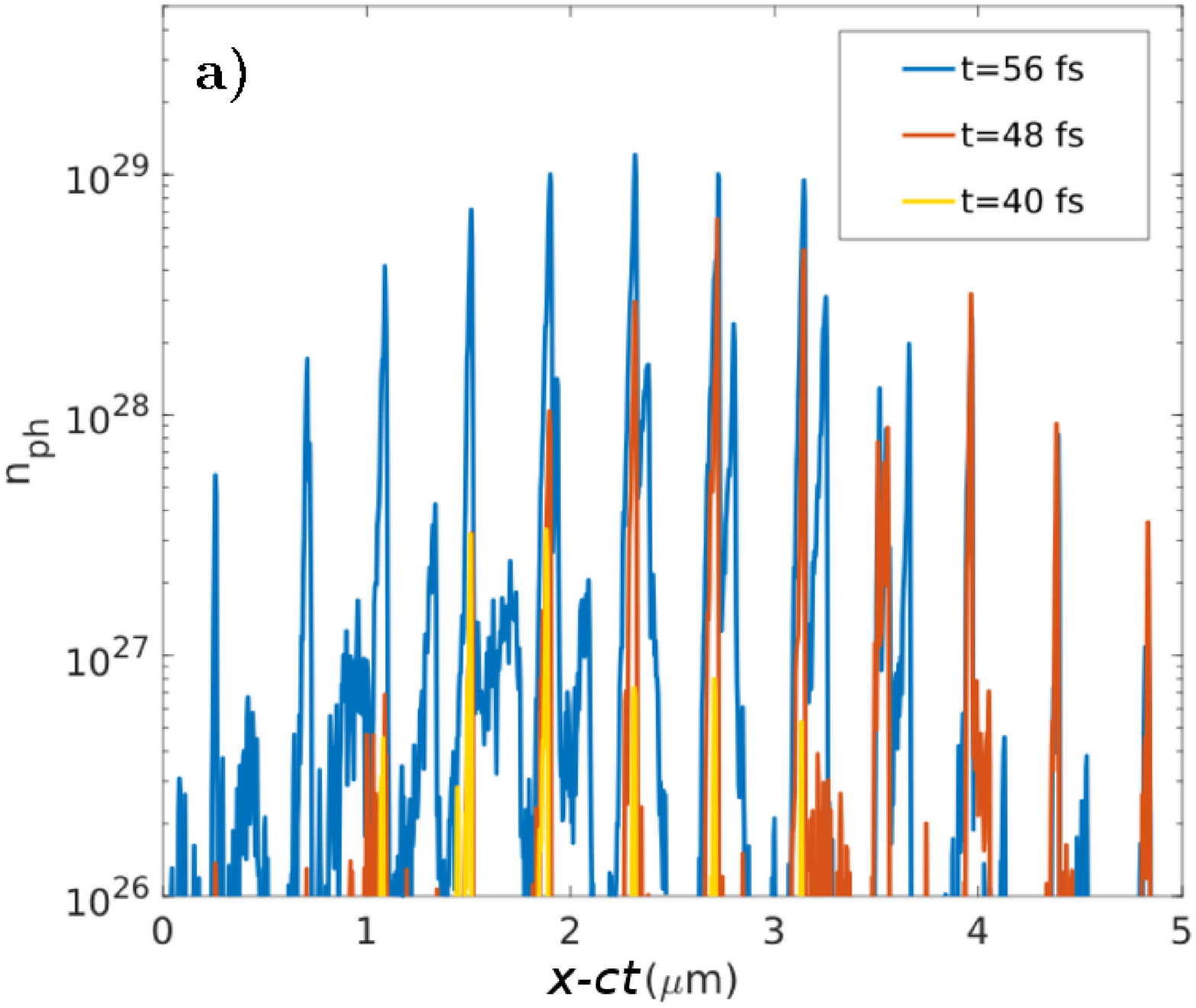}
\includegraphics[width=54mm]{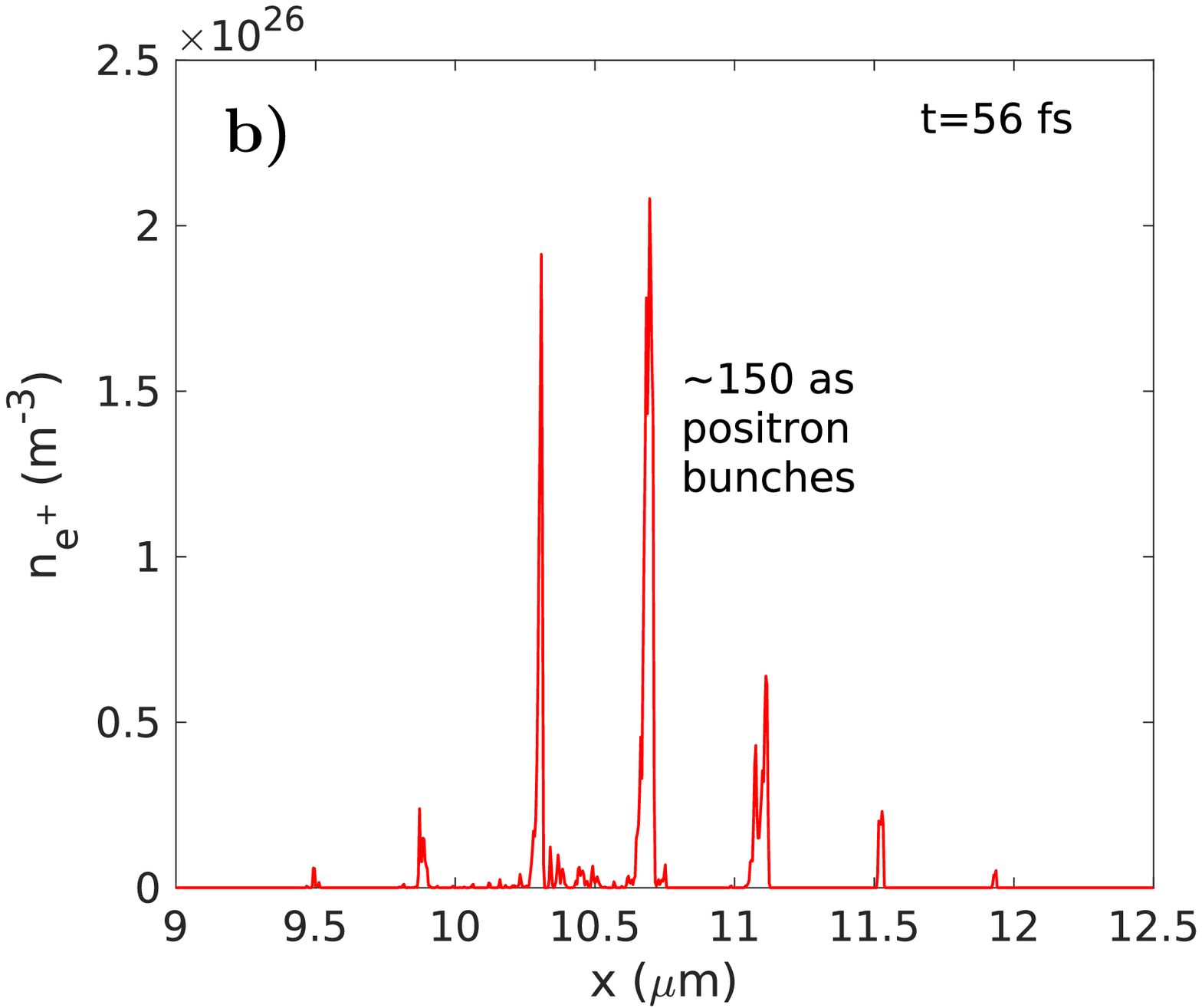}
\includegraphics[width=54mm]{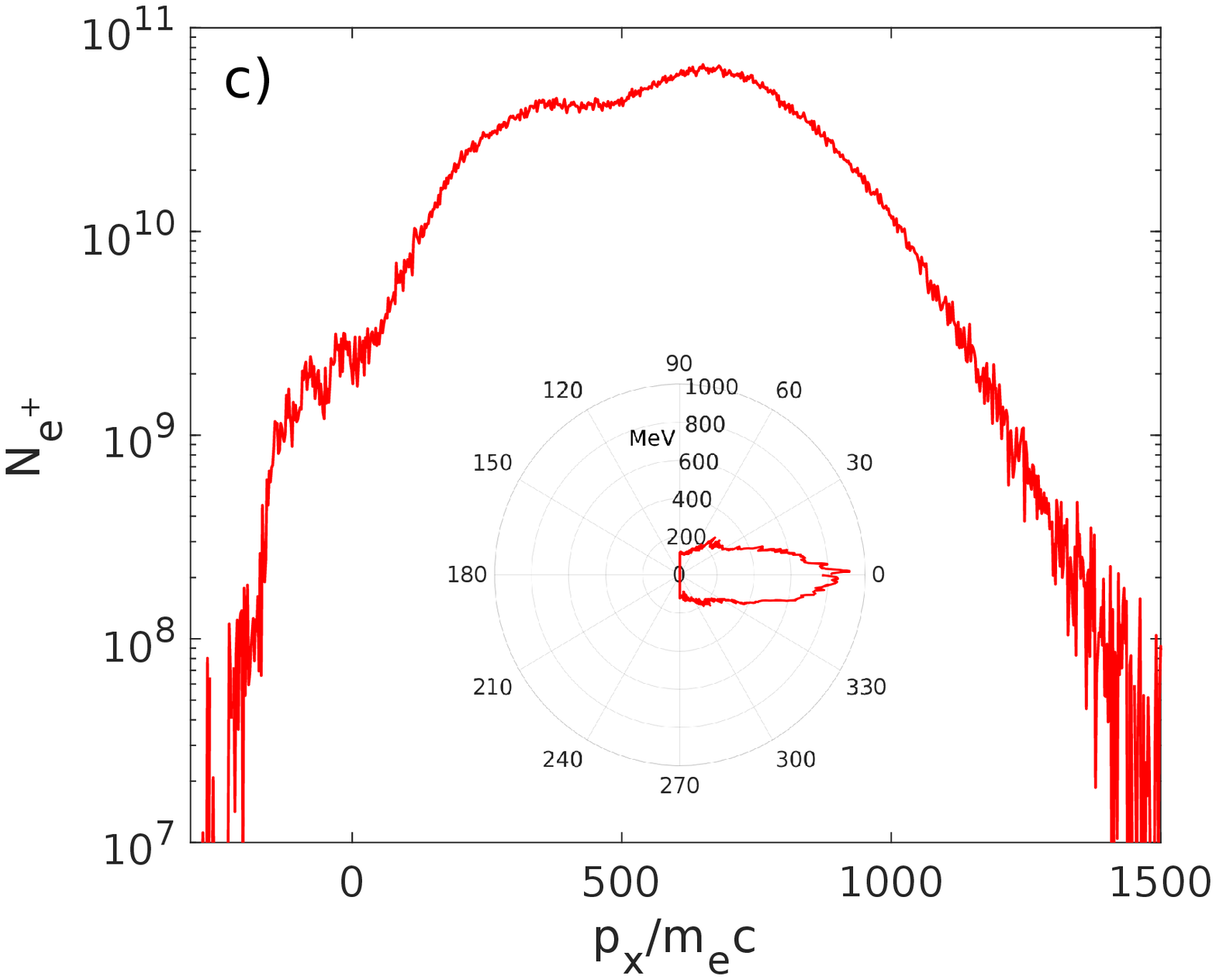}

\caption{(a) Evolution of photon density along the line $y=0$. At later time beside the primary peaks also secondary peaks appear. (b) Positron density along the $y=0$ axis. (c) Longitudinal momentum spectrum of positrons and their angular distribution shown in the inset.}
\label{posAtto}
\end{figure*}

Fig. \ref{interaction2} shows the distributions of energetic electrons and gamma photons. Thin nano-layer of electrons are formed near the positions where the ponderomotive force is maximal ($\xi_0$, which corresponds to the phase $\pi/4$ of the laser wave) and propagate together with these nodes with velocity close to the phase velocity. Very often the high electron energy
observed in such targets is attributed to the betatron resonance between the laser field and electron oscillation around the nanorods (or nano-wires) \cite{betatronRes, RPA_MLF}. The large $D_{sp}$ considered here means that the static self-fields of nanorods are negligible and can be omitted in the vacuum region. The electrons co-propagate with the laser field and thus experience a constant accelerating force. 

%The final energy gain can be calculated from the product of the ponderomotive force and the traversed distance:

%\begin{equation}\label{eq:energy}
%\epsilon_e^f=m_e c^2  \int_0^{L_r} \frac{a_0^2}{4\gamma(x)} \frac{\partial f(\xi_0,t)^2}{\partial \xi} \mathrm{d}x,
%\end{equation}
%where $f(\xi,t) $ is the function describing the laser wave-form, with $\xi=x-ct$, and $L_r$ is the length of nanorod.

During the laser propagation, the electrons are also accelerated in transverse direction and eventually reach the neighbor nanorod. This transversal velocity component means that the electrons in $x$ direction are slower than the laser wave (their velocity is smaller than the phase velocity) and fall behind the original position where they have been extracted from (in the moving frame of the laser pulse). In this way, electrons propagating farther away from the nanorod enter the region where the ponderomotive force has opposite sign and they get decelerated. A single electron bunch extracted by the negative electric field (blue), shown in Fig. \ref{interaction2}, moves upwards and at some distance the ponderomotive force changes its sign and slows
it down. These electrons start to lose energy up to the point when they reach the laser phase where the ponderomotive force is positive again and are subsequently re-accelerated in the forward direction. However, the electric field now has the opposite sign. In Fig. \ref{interaction2}a,  the energetic photons are emitted by the electrons which already reach and cross the neighbor nanorod. The opposite sign of the electric field means that the change in transversal momentum is large which results in more intense radiation, i.e. more photons. The electrons propagating between the nanorods in the vacuum region have large longitudinal momentum and are continuously accelerated. These electrons are basically the same as those presented in Ref. \cite{microWire} and emit large amount of photons when the pulse gets reflected from the plasma slab. The population of high energy photons ($>300$ MeV) finally will be higher between the nanorods in the high field region, as it is shown in Fig. \ref{interaction2}c and in Fig. \ref{turnBack}a.

As the pulse propagates between the nanorods, more electrons reach the neighboring rod and will eventually turn back. This is clearly visible in Fig. \ref{turnBack}a. These electrons are also well compressed by the positive ponderomotive force and emit photons in the form of short bunches. However, their energy and coherency is lower than photons emitted between nanorods because of the intermediate deceleration experienced during the turning back. The divergence of high energy photons is very small, as shown in Fig. \ref{turnBack}b. 

The evolution of gamma photon atto-bunches is presented in Fig. \ref{posAtto}a. Initially very short ($< 100$ nm) bunches of photons are generated near the laser propagation axis, but later the photons emitted by the returned electrons also appear, which can be recognized from the secondary peaks of the blue line in Fig. \ref{posAtto}a. The high energy photons can decay to electron-positron pairs by interacting with the reflected field. The density of positrons along the axis is shown in Fig. \ref{posAtto}b which confirms that using a single laser pulse and a nanorod array target positron attobunches can be produced similar to those reported in Ref. \cite{microWire}. In this case, approximately 0.1 \% of the energetic photons is converted into matter. Such a positron yield can not be achieved with a simple flat foil target at this laser intensity. In this interaction, after reflection from the flat foil, about 90 \% of laser energy is absorbed and 60 \% of the absorbed energy is transferred to the ions via Coulomb explosion. At $t=56$ fs, 37 \% of the energy is in the plasma electrons and the rest is in photons and pairs. The resulting energy conversion efficiency from laser pulse to photons is about 3 \% and to positrons is $\sim 3\times 10^{-5}$. Fig. \ref{posAtto}c shows the momentum distribution of the generated positrons and in the inset angular distribution is shown. The maximum energy is close to 1 GeV and the positrons are well collimated in a narrow cone angle of $\sim 10$ degrees, as expected from the small divergence of the decaying photons. 

\begin{figure}[h]
\centering
\includegraphics[width=0.23\textwidth, trim= 20mm 0 10mm 0]{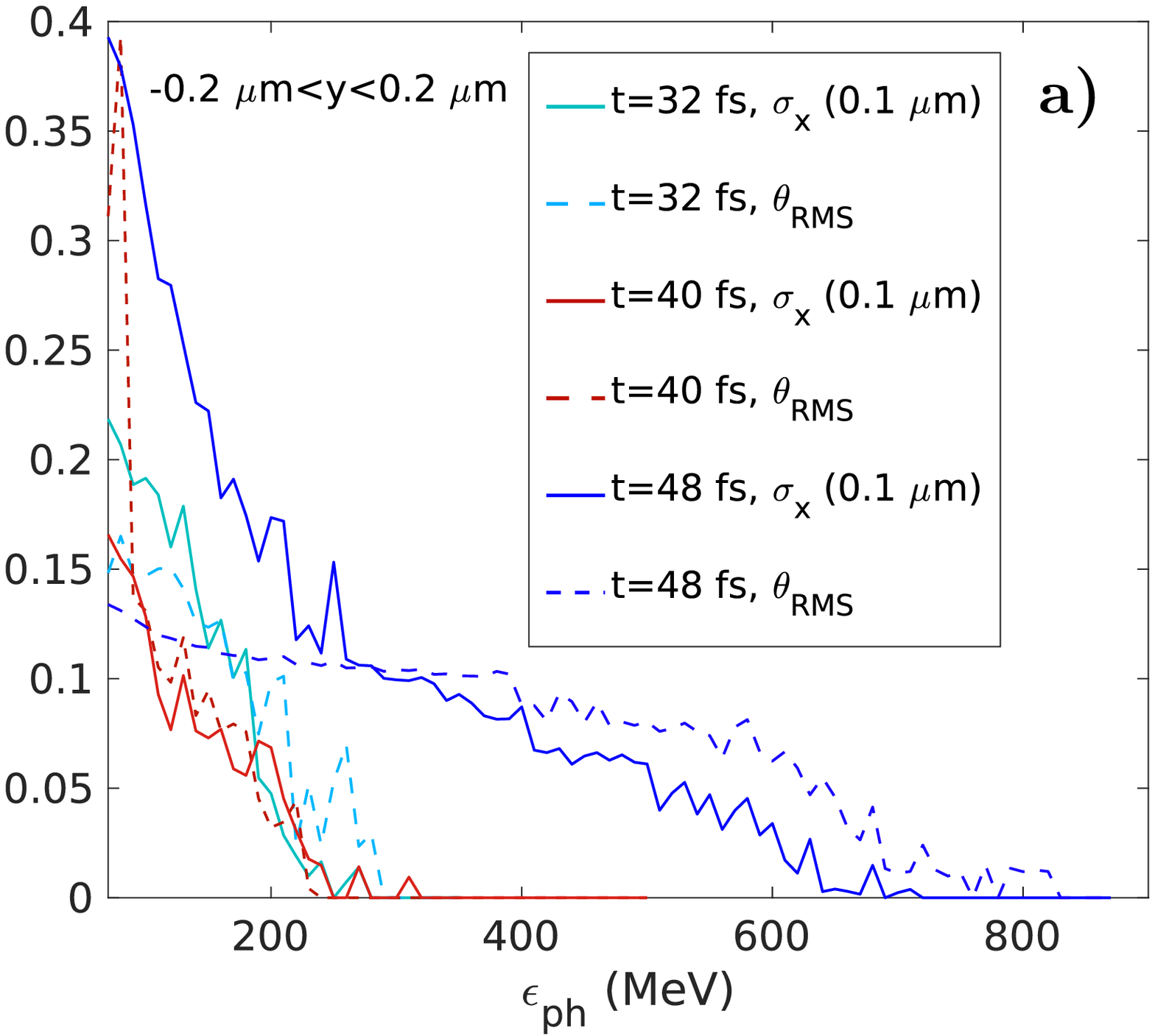}
\includegraphics[width=0.24\textwidth, trim= 20mm 0 10mm 0]{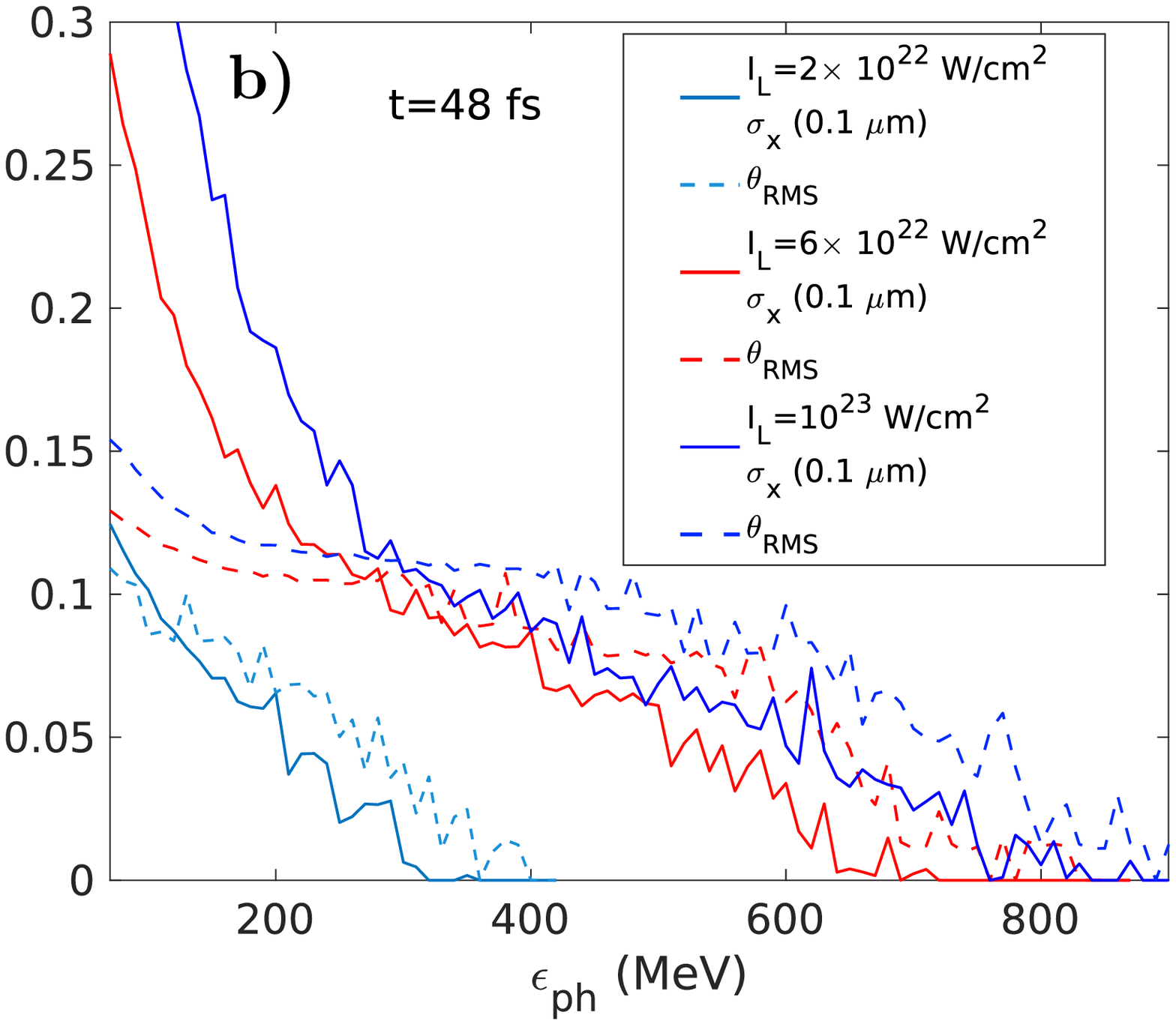}

\caption{ (a) Bunching factor (in units of 0.1 $\mu$m, full lines) and RMS angle (in radian, dashed lines) of photons versus energy at different time instances. (b) The same as in (a), but for different laser intensities at the given time instance. }
\label{intensityVar}
\end{figure}

The statistical properties of photons are further analyzed in Fig. \ref{intensityVar}a, where the standard deviation of photon positions is calculated. This can be interpreted as a bunching factor characterizing the spatial distribution of emitted photons: $\sigma_x=\sqrt{\Sigma (x_i-<x>)^2/N}$, where $x_i$ is the position of individual photons and $N$ is their total number in one bunch. This value is calculated for several bunches and the average is plotted as a function of energy. Similarly the root mean square angle ($\theta_{RMS}$, where $\theta=\arctan \frac{p_y}{p_x}$) is also calculated in each energy bin of the spectrum. Both values decrease with increasing photon energy, which means that the more energetic gamma photons are more collimated and have greater coherency. The photon bunch length, after filtering out the low energy photons, is on the order of 10 nm and their angular spread is about $6^\circ$. 

The nano-bunching of electrons is limited by the grid size used in the simulation because of numerical dispersion effects. We have found that with this grid resolution the result converges, but in the case of higher laser intensities, where the stronger ponderomotive compression would result in even thinner and denser electron bunches, smaller grid cells are necessary. By using the same grid resolution the photon coherency should remain the same or higher when higher laser intensity is applied. However, Fig. \ref{intensityVar}b shows that this is not the case and the higher intensity results in wider and less coherent photon bunch. This could be attributed to the radiation reaction effect, where the electrons lose momentum and energy after each emission of a single photon. The recoil effect manifests itself also in the larger angular spread and in the similar cut-off energy for two intensity vales. Despite the stronger fields, the electrons can not gain much higher energy 
due to more frequent photon emission which takes away energy. 

The continuous acceleration of electrons means that the emitted radiation power scales strongly with the length of nanorods ($L_r$) \cite{nanoforest}. The photon energy obtained using shorter rods and higher intensity can be reproduced using longer nanorods and lower laser intensity. However, coherence is not guaranteed over a longer propagation distance and in practice nanorods with such length ($> 10\,\mu$m) are difficult to produce. Instead a stack of nano-plates is more appropriate to consider or a single slit or hole on an over-dense target could be also used \cite{attobunches, sub_femtoXray}. The 2D geometry used here does not allow electrons to bypass the nanorods and they always cross them, which is not completely realistic. Therefore the simulations presented here describe more realistically the interaction of P-polarized laser pulses with micro-gaps which are separated by $\sim 100$ nm thick walls and provide insight into the electron dynamics during the laser propagation. A full 3D simulation would show us lower efficiency in number of emitted photons and the divergence in the $z$ direction would be also revealed. 

In conclusion, it has been shown that attosecond physics is not limited to photon energy around 100 eV but can be extended to the area of quantum electro-dynamics where attopulses consisting of gamma photons of energy above 100 MeV are generated. Numerical simulations have confirmed that very short electron bunches can be accelerated to GeV energies by using nanorod array targets and laser intensities below $10^{23}$ W/cm$^2$. These bunches emit gamma photons with the similar spatial distribution. At higher intensity, the emission is less coherent due to the radiation reaction effect. The length of the generated photon bunches is below 100 nm and propagates with $<10$ degrees divergence. These unique photon bunches can be used in diagnostics of high energy density matter or in nuclear physics experiments. We have also shown that by allowing the collision of the gamma photons with the same laser pulse reflected from a flat foil results in the creation of attosecond positron bunches. 

\begin{acknowledgments}
The ELI-ALPS project (GOP-1.1.1.-12/B-2012-0001, GINOP-2.3.6-15-2015-00001) is supported by the European Union and co-financed by the European Regional Development Fund.
\end{acknowledgments}

\bibliography{Manuscript}

\end{document}